\documentclass[11pt,twoside]{article}

\usepackage{asp2006}
\usepackage{epsf}
\usepackage{lscape}

\newcommand{\ee}[1]{\mbox{${} \times 10^{#1}$}}
\newcommand{\eten}[1]{\mbox{$10^{#1}$}}
\newcommand{\lir}{\mbox{$L_{\rm IR}$}} 
\newcommand{\lmol}{\mbox{$L_{\rm Mol}$}} 
\newcommand{\mvir}{\mbox{$M_{\rm vir}$}} 
\newcommand{\lsun}{\mbox{L$_\odot$}}
\newcommand{\msun}{\mbox{M$_\odot$}}
\newcommand{\sfr }{\mbox{$\dot M_{\star}$}}
\newcommand{\msunyr}{\mbox{M$_\odot$ yr$^{-1}$}}
\newcommand\cmv{\mbox{cm$^{-3}$}}
\newcommand{\av}{\mbox{$A_V$}} 
\newcommand{\lhcn}{\mbox{$L_{\rm HCN}$}}
\newcommand{\lco}{\mbox{$L_{\rm CO}$}}

\newcommand{\mdense}{\mbox{$M(dense)$}}
\newcommand{\mgas}{\mbox{$M(gas)$}}
\newcommand{\kms}{\mbox{km s$^{-1}$}}
\newcommand{\spitzer}{\mbox{\it Spitzer}}
\newcommand{\epsstar}{\mbox{$\epsilon_\star$}} 
\newcommand{\speed}{\mbox{``$\epsilon$''}} 
\newcommand{\sfrff}{\mbox{$SFR_{ff}$}} 
\newcommand{\tdep}{\mbox{$t_{dep}$}} 
\newcommand{\hcop}{HCO$^+$}

\begin{document}
\title{Star Formation in Molecular Clouds?}   
\author{Neal J. Evans II}   
\affil{The University of Texas at Austin, Department of Astronomy, 1 University
Station, C1400, Austin, TX 78712-0259, USA}    

\begin{abstract} 
Using studies of nearby star formation with Spitzer, I will argue
that star formation is restricted to dense cores within molecular clouds.
The nature of these dense cores and their connection to star formation
will be discussed. Their distribution over masses and over the cloud
is similar to that of stars, and their efficiency of forming stars
is much higher than that of the whole cloud. Moving to regions forming
more massive stars, we find that the mass distribution of the dense
clumps is similar to that of OB associations. The infrared luminosity
per unit mass of dense gas is high and comparable to that seen in starburst
galaxies. The relation between star formation and dense gas appears to be
linear. Understanding the Kennicutt-Schmidt law requires an understanding
of what controls the conversion of gas into the dense entities where
stars actually form.

\end{abstract}


\section{Introduction}   

The image that comes to mind when we talk about star formation across
disciplinary lines is that of the ``sight-challenged" villagers and the elephant.
Each feels a different part of the elephant and comes to different conclusions
about what it is. Discussions of star formation within the Milky Way use language
unfamiliar to those who work on star formation in other galaxies and vice versa. If
we are to progress, we need to find a common language.

Studies of low mass star formation in nearby (typically within 500 pc) clouds
can speak in detail about the distribution of star formation, the timescales
for various phases, and the current efficiency in stellar mass per cloud mass.
Detailed, predictive theoretical models exist for the formation of individual
stars (e.g., \citealt{shu1987}).
There is a strong focus on detailed studies of the flow of material from envelope
to disk to star and on the connection to planet formation (e.g., many articles
in \citealt{ppv}). We can also provide observational constraints on the origin
of the initial mass function. However, these studies are not directly relevant
to star formation on galactic scales because low-mass star formation is not
observable at large distances.

Regions forming more massive stars begin to show up at about 440 pc 
\citep{hirota2007} in the Orion complex
but most lie at distances of kpc, with the most impressive being in the 
Galactic ring, located about 5 kpc from the center of the Milky Way
\citep{clemens1988}. 
For the most part, we are really talking about formation of clusters of
stars, including both high and low mass stars, though most forming clusters do
not survive \citep{lada2003}.
Our knowledge of these regions is much less detailed, owing to the lack of spatial
resolution, and theoretical predictions are less well developed. Considerable
theoretical progress is being made, but there are fundamental disagreements
on some issues (cf. \citealt{krumholz2005}, \citealt{bonnell2006},
\citealt{martel2006}).
For most of the distant regions, we do not separate individual stars, but
rather we characterize the star formation in terms of surrogates, such as 
the far-infrared luminosity (\lir) and the mass of molecular material.
While less detailed, these crude measures are more compatible with what
can be done in other galaxies.

Discussion of star formation in other galaxies speaks nearly exclusively in 
terms of ``Schmidt Laws". The original version compared the scale heights of
young stars and the HI gas and concluded that the star formation rate is
proportional to the square of the {\it volume} density (Schmidt 1959). Were this to be
done today, recognizing that star formation is restricted to molecular gas,
one would probably infer a linear relationship.  However, {\it surface} density
is more amenable to study in other galaxies and modern versions find relations between
the surface densities of star formation and gas \citep{kennicutt1998}. 
A power law above a threshold provides a remarkably good fit:
$$\Sigma(SF)({\rm \msun\ yr^{-1} kpc^{-2}}) = 
(2.5\pm0.7) \times 10^{-4} (\Sigma(gas)/1 \msun {\rm pc^{-2}})^{1.4\pm0.15}$$
over a wide range of galaxy types and star formation rates.
Several explanations have been offered for the value of the exponent
(\citealt{elmegreen2004}, \citealt{krumholzmckee2005}, \citealt{shu2007}). 
Models of galaxy formation and evolution generally use a Kennicutt-Schmidt relation
to simulate star formation, which is treated as ``sub-grid physics."

To return to the elephant analogy, each villager feels a different part of
the elephant, but also speaks a different language, which is only partly
understood, and often misinterpreted, by the others. But the need for 
communication is great as rapid progress is occurring in studies of massive
star formation in our galaxy. At the same time, studies of galaxy formation 
are progressing rapidly, including both
archaeological studies of the oldest stars and direct look-back studies of
high-z luminous starbursts. Can we find a common language?

I will summarize recent progress in studies of both local, low-mass
and clustered, high-mass star formation in our Galaxy, and then I will
discuss some connections that can be made.

\section{Low-mass Star Formation in Nearby Clouds}

For convenience, I will focus here on recent results from the {\it Spitzer}
legacy project, ``From Molecular Cores to Planet-Forming Disks", known
as ``Cores to Disks" or c2d (Evans 2003). In particular, I will concentrate on
studies of five large nearby clouds. For three of these clouds (Perseus, Serpens,
and Ophiuchus), we have complementary information on the dust emission at 1 mm
using Bolocam (\citealt{enoch2006}, \citealt{young2006}, \citealt{enoch2007})
and molecular line emission \citep{ridge2006}. These studies provide the first
complete coverage of molecular clouds in tracers of gas, dense cores, and young
stellar (or substellar) objects (YSOs) down to luminosities of about 0.01 \lsun.

Comparison of the CO and $^{13}$CO maps with the Bolocam maps shows that the dense
($n > 2\ee4$ \cmv ) cores traced by dust continuum emission at millimeter
wavelengths are distributed very non-uniformly over the cloud. Dense cores are highly
clustered, and many lie along filaments traced by $^{13}$CO. The dense cores
strongly prefer regions of high overall extinction, with 75\% of dense cores found
within extinction contours that contain only 0.1 to 0.3 of the total cloud mass
\citep{enoch2007}.
Comparison of the dense core distribution to that of YSOs shows a strong correlation,
especially for the early stages of star formation. The dense cores are clearly
the sites of star formation; most of the cloud is completely inactive.

Until spectral types are available for more of the YSOs, we cannot construct IMFs
for the clouds, but all data so far are consistent with a typical IMF. The core
mass distribution for {\it starless} cores, averaging over all 3 clouds, is
remarkably consistent with the usual IMF, but shifted to larger masses by a 
factor $<4$,
suggesting that the IMF is set by the core mass distribution with
an efficiency $>25$\% (Enoch et al. in prep.). Similar conclusions have
been reached by \citet{alves2007} 
using extinction mapping to trace cores; their method
produces less uncertain masses but probes lower densities on average. Of course,
the similarity of the IMF and the core mass distribution could be fortuitous if
cores of different masses have different lifetimes \citep{clark2007}.

We define the actual efficiency of star formation as the fraction of the mass
that ends up in stars (or substellar objects):
$$\epsstar = M_{\star}/(M_{\star}+M_{\rm gas}) \approx M_{\star}/M_{\rm gas}. $$
The {\it final} efficiency, after the gas is consumed or dissipated, 
is impossible to determine
directly since the gas lifetime vastly exceeds human lifetimes, but we can constrain it.
In practice we measure the {\it current} efficiency using the current mass in
stars and in gas.
To get a star formation rate (\sfr), we assume a mean stellar mass of 0.5 \msun\
(consistent with actual determinations in one cloud, Alcal{\`a} et al.
in prep.), and a timescale
of 2 Myr (essentially the half-life of infrared excesses that the c2d observations
are sensitive to). 
It is possible for these nearby regions to separate the efficiency, as defined
above, from the question of the {\it speed} of star formation.
We define the timescale for gas depletion, as in extragalactic studies:
$$\tdep = \mgas/\sfr,$$
and the speed is $1/\tdep$.
\citet{krumholztan2007} define the speed as
$$ SFR_{\rm ff} = t_{\rm ff}/t_{\rm dep} = \sfr t_{\rm ff}/\mgas. $$
This is the speed ($1/\tdep$) normalized to the maximum likely speed ($1/t_{\rm ff}$),
or the star formation rate normalized to the maximum possible in a free fall time.

The preliminary results for these quantities for the five large clouds studied by c2d
are summarized here.  The total mass in YSOs 
is about 2\% to 4\% of the cloud mass (e.g., \citealt{harvey2007}, Evans et al. in prep.) 
consistent with earlier conclusions that star formation is inefficient. 
The total mass in dense cores is only 1\% to 4\% of the total cloud mass included
in the $\av = 2$ contour. Low star formation efficiency begins with inefficient 
core formation.
The values of \sfr\ range from 6.0 to 71 \msun/Myr, a significant range. 
Star formation is slow: the depletion timescales for the whole molecular clouds are
$\tdep = 50$--90 Myr, much longer than most estimates of cloud lifetimes of 5--10 Myr. 
While further star formation is likely, the final efficiencies 
are likely to remain low. In contrast, the depletion times for dense cores, the actual
sites of star formation, are 0.6--3.5 Myr, and masses in dense cores are similar to
current masses in YSOs. Star formation is fast and efficient {\it in dense gas}.

What about the speed relative to that obtained from assuming free fall?
In calculating \sfrff, we use the mean density of the relevant unit to 
calculate the free fall time. The \sfrff\ is 0.02 to 0.03 for the cloud as a whole,
consistent with long-standing arguments against efficient star formation
at the free-fall rate (e.g., \citealt{zuckerman1974}).
Considering the clump that is forming a cluster, the \sfrff\ increases to 0.1 to 
0.3, but if we use the mean density of an individual dense core, the \sfrff\ is
only 0.03 to 0.12. Star formation remains ``slow" compared to a free fall time.
This slowness cannot be blamed entirely on a slow prestellar phase; comparison
of the numbers of dense, starless (prestellar) cores to the numbers of
protostellar cores yields lifetimes of a few free-fall times once the density
exceeds about $2\ee4$ \cmv\ (Enoch et al. in prep).

Does the surface density relation \citep{kennicutt1998} have any relevance
to local star formation? This one is
hard to test locally because of incompleteness, but we can make a rough
estimate. With a local gas surface density of 6.5 \msun\ pc$^{-2}$ (L. Blitz,
pers. comm.), of which about 85\% is HI (\citealt{levine2006}, \citealt{dame1993}), 
the Kennicutt relation would predict 
$\Sigma(SF) = 3.4\ee{-3}$ \msun\ yr$^{-1}$ kpc$^{-2}$. 
\citet{lada2003} estimate that embedded clusters contribute, to within a factor of 3,
$\Sigma(SF) = 3\ee{-3}$ \msun\ yr$^{-1}$ kpc$^{-2}$ within a radius of 0.5 kpc.
This is surely an underestimate until we have a more complete census of all
clouds within the local area, but it is already remarkably close to the
prediction. A more extensive study of this topic can be found in 
\citet{blitz2006}. It is interesting to note that the low-mass star formation in
the clouds studied by the c2d project would not be apparent to the usual
tracers used in extragalactic studies, with the possible exception of \lir.

The lessons we should take from these very detailed studies of nearby clouds
are as follows. The fundamental units of star formation are dense cores, not
molecular clouds, per se (hence the question mark in the title). The cores are
not located randomly over cloud faces, but are concentrated in clumps, which
are forming clusters, and filaments, which tend to form smaller aggregates. The
core mass distribution {\it may} determine the IMF. Star formation in molecular
clouds is very inefficient (2--4\%), but quite efficient in dense cores ($>25$\%).
Star formation is slow in terms of a free-fall time, especially for the cloud as 
a whole, but somewhat faster in terms of a free-fall time (and much faster in 
absolute terms) in dense cores.

\section{Massive, Clustered Star Formation}

With the exception of a few, relatively nearby regions of massive, clustered
star formation, such as the Orion region (e.g., \citealt{hillenbrand1997}, 
\citealt{lada2003}, \citealt{allen2007}), we do not
have such detailed information on stellar IMF, ages, efficiencies, etc.
Instead we accept cruder measures, but these provide a bridge to star formation
in other galaxies. Theories are not well developed yet, but progress is
being made, as discussed at the meeting by Bonnell, Dobbs, and Krumholz.

There have been a number of surveys for dense gas, but they have all been
directed toward sign-posts of star formation, such as water masers \citep{plume1992} 
or {\it IRAS} sources (\citealt{sridharen2002}; \citealt{beuther2002}). 
An unbiased survey of 1 mm continuum
emission now underway using Bolocam will provide a more complete census
\citep{williams2007} and deeper surveys with SCUBA-2 and far-infrared
surveys with Herschel will follow. We should soon have a much more complete
picture of star formation sites in the Milky Way.

For now, I will focus on results from the survey toward water masers \citep{plume1992}, 
which has been followed up by many other studies obtaining gas densities from
multitransition CS observations \citep{plume1997}, images of dust emission at 350 \micron\
\citep{mueller2002}, maps of CS $J = 5-4$ emission \citep{shirley2003}, and maps
of HCN emission \citep{wu2005}. While these references call the objects
``dense cores", I will adopt the convention of McKee and co-workers
(e.g., \citealt{krumholztan2007}) and refer to them
as dense {\it clumps}. The dense clumps are the sites of cluster formation and
we reserve the term core to refer to the structure that forms one or a few stars.

The dense clumps can be fitted to power law density profiles with exponents very 
similar to
those of dense cores forming low mass stars, but the density, measured at the same
radius is typically 100 times higher and the linewidths are typically 16
times wider. Some show evidence in the line profiles for inward motions \citep{wu2003},
similar to those seen in some low mass cores. In general, they
are scaled up versions of low-mass dense cores. This point is important, as some
authors use ``clump" to mean something less dense than a core. Massive, dense
clumps are {\it denser} than the low mass cores. The clumps within them are
presumably even denser, but they have been hard to separate cleanly.

The mass function of the dense clumps is steeper that that of molecular clouds
or of less dense clumps traced by CO isotopes (Shirley et al. 2003), but it is
incomplete below about 1000 \msun. Above that level, it is similar to the
distribution of the
total masses of OB associations (\citealt{massey1995}; \citealt{mckee1997}).
These dense clumps are very likely to be the sites where clusters and OB 
associations form. However, it has been difficult to study the forming stars
themselves because of the heavy extinction and large distances. The GLIMPSE
legacy project \citep{benjamin2003}
using the IRAC instrument on {\it Spitzer} is revealing
clusters in some of these, but the strong diffuse emission in the IRAC bands
makes it difficult to extract detailed information on the stellar content
(Nordhaus et al. in prep).

Our observational measures are the luminosity in a molecular line or in 
dust continuum at long wavelengths, which are tracers of the mass of dense gas,
and the bolometric luminosity measured by integrating over the full spectrum
of dust emission. The line luminosities of tracers of dense gas, in particular
CS and HCN, trace very well the virial mass of dense gas (e.g., Wu et al.
in prep.).
The integrated infrared luminosity (\lir) traces the star formation rate (\sfr)
given enough time for the IMF to be reasonably sampled \citep{krumholztan2007}.
Both these measures are subject to fluctuations about mean values of at least
a factor of 3. Despite the variations, \lir\ correlates well with \lmol\ or
\mvir. 

Without a detailed census of the stellar content, we cannot compute the star formation 
efficiency in the same sense ($\epsstar$) as we could for low mass star formation,
but the similarity of the mass function to that of OB associations suggests
reasonably high efficiency in the dense gas. 

We will use an efficiency measure common in the extragalactic
context as the star formation rate per unit mass of gas 
($\speed = \sfr/\mgas$. Note that this ``efficiency" is really the
speed ($1/\tdep$) unnormalized to the free fall time.
In observational terms, this is measured by $\lir /\lmol$.
As for low mass star formation, the \speed\ is very low for molecular clouds
as a whole, but much higher for dense gas  (e.g., a factor of 30 higher
in one study, \citealt{mueller2002}).
\citet{krumholztan2007} have argued that star formation is
also slow relative to a free-fall time (\sfrff), though again there is
some evidence that \sfrff\ is higher in the densest gas, probed for example by
the CS $J = 5-4$ transition. 

These massive dense cores provide the connection point between detailed
studies of star formation in the Milky Way and extragalactic star formation.

\section{Star Formation in Galaxies}

Since we seek a connection between studies of star formation in our
Galaxy and extragalactic star formation, we will focus on studies using common
tracers. In dusty galaxies, the star formation rate is traced by \lir. This is
basically a calorimetric method: the dust absorbs all the energy from young stars
and re-radiates in the infrared. The advantage is that there is no need for 
uncertain extinction corrections, which are needed for ultraviolet or H$\alpha$
measures of star formation. However, one is assuming that {\it all} the light
is indeed absorbed, so some star formation can be missed (this problem can
apparently be alleviated by combining infrared and H$\alpha$ observations, as
described by Calzetti at this meeting). Also, a calorimeter measures any heat
input, so heating by older stars or an AGN can confuse matters. In practice,
this method is applied only to systems where star formation overwhelms the heating
by older stars. Assessing the contribution of AGNs can be trickier, especially for
high-z systems. A common calibration, based on a continuous burst model is that
$$ \sfr (\msunyr) = 1.7\ee{-10} \lir(\lsun)$$
\citep{kennicutt1998}, with an averaging time of 10 to 100 Myr. This number
may vary by a factor of at least 2 depending on the star formation
rate and age of a starburst (see \citealt{krumholztan2007}).

How does \lir\ correlate with various tracers of gas? If CO is used to trace
the molecular gas, the relation is strongly non-linear. 
For CO, the \speed\ increases by a factor of 100 as \lir\ rises from
\eten{10} to \eten{14} \lsun\ \citep{Solomon and Vanden Bout (2005)}.
Even higher \speed\ may be observed in high-z submillimeter galaxies
\citep{greve2005}.

If the \lir\ versus \lmol\ plot is made using HCN $J=1-0$, instead of CO,
the correlation is better, the scatter is less, and the relation is {\it linear}
(\citealt{gao2004a}, \citealt{gao2004b}). In this case, the \speed\ is constant over
more than 3 orders of magnitude in \lmol. The HCN line is sensitive to 
relatively dense gas, which reminds us that it was only the dense molecular
gas that is involved in star formation in well-studied local molecular clouds.

\begin{figure}
\plotone{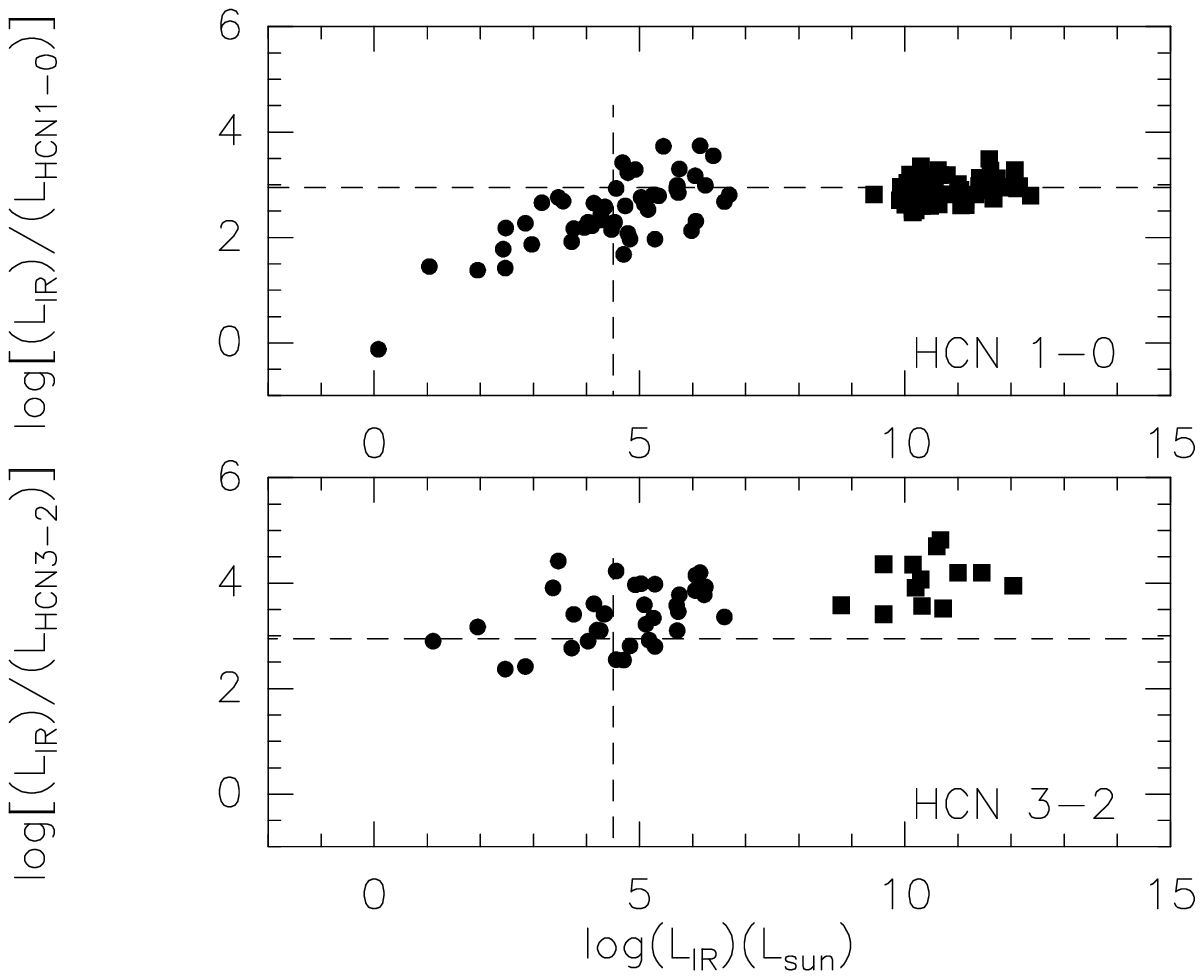}
\caption{
Correlations between the distance-independent ratio \lir/\lmol\
vs. \lir\ for HCN $J = 1-0$ amd HCN $J = 3-2$. The squares are HCN $J = 1-0$
observations of galaxies \citep{gao2004a} in the first panel,
and HCN 3-2 observations of galaxies \citep{paglione1997} 
in the second panel. The circles are the equivalent data for massive
dense clumps in our Galaxy. The horizontal dashed line in the top plot indicates
the average \lir/\lhcn\ ratio for galaxies; the vertical
dashed line in the top plot shows the cutoff at $\lir =10^{4.5}\lsun$.
These two lines are also shown in the bottom plot to indicate the relative
shifts of \lir/\lmol\ between HCN $J = 1-0$ and HCN $J = 3-2$.
\label{lirfig}}
\end{figure}

In order to compare directly the situation in local clouds to that in starburst
galaxies, Wu et al. (2005) surveyed the HCN $J=1-0$ line in Galactic clumps.
The results showed that, above a threshold in \lmol\ or \lir, the Galactic clumps
fit a linear relation between \lir\ and \lmol, essentially identical to that of
the starburst galaxies. Above the threshold in \lir\ of about \eten{4.5} \lsun,
the ``efficiency," \lir/\lmol, is very similar in the clumps to that in starbursts
(Fig. \ref{lirfig}).
Below the threshold, \lir/\lmol\ drops rapidly with decreasing \lir.
While data on the HCN $J=3-2$ line are scarce for galaxies, the currently
available data tell a similar story, but with a higher ratio of \lir/\lmol\
(Fig. \ref{lirfig}). A higher ratio is not surprising because the HCN $J = 3-2$
line traces still denser gas.

HCN is not the only dense gas tracer. Many other tracers, including CS and
\hcop\ have been used in studies of dense gas in our Galaxy.
\citet{carpio2006} recently suggested that \hcop\ could be a better 
tracer in galaxies, but \citet{papadopoulos2007} came to the opposite
conclusion and noted the need for more observations of higher J transitions
to constrain the conditions in the dense gas. A recent multitransition study 
of CS, HCN, and \hcop\ in  two local ULIRGs \citep{greve2006} found many
similarities to the properties of dense clumps in the Galaxy. They concluded
that, for the transitions they studied, \hcop, HCN, and CS probe progressively
denser gas.

\section{An Emerging Picture}

The comparisons in the last section suggest that local studies of massive
star formation can indeed provide insights into starbursts. Qualitatively,
a modest starburst can happen if a galaxy has a lot of molecular gas, with 
some fraction of it being dense. For an extreme starburst, a large fraction of 
the molecular gas must be dense. At the extreme, the entire ISM might look like
the dense, cluster-forming clumps in our Galaxy.

Quantitatively, we offer some ``Schmidt laws" (Wu et al. (2005):

$$ \sfr (\msunyr) = 1.4\ee{-7} \lhcn (\rm{K\ \kms\ pc^2}) $$

$$ \sfr (\msunyr) = 1.2\ee{-8} \mdense (\msun) $$
where \lhcn\ is the monochromatic line luminosity of the HCN $J=1-0$ line
in observer's units,
and we have assumed that $\sfr = 2.0\ee{-10} \lir$ and $\mdense (\msun) = 7  \lhcn$
(K \kms\ pc$^{-2}$) \citep{wu2005}.
While these are not yet of practical use, {\it CARMA} and {\it ALMA} will be supplying
much more observational data on the lines in the coming years.
In the meantime, we should try to understand the relationship between these
{\it linear} relations for dense gas and the non-linear relations for less dense
gas.

First, we should ask why \lir/\lhcn\ is constant from $\lir = \eten{4.5}$ to
\eten{13} \lsun, but drops sharply for lower \lir. At first glance, the constancy
is quite puzzling for Galactic clumps even if the star formation rate is linear
with the amount of dense gas because one would expect the mass of the most massive
star to increase with the number of stars formed and the stellar luminosity is
a non-linear function of the stellar mass. Thus, one might expect \lir\ to
increase non-linearly with \mdense. In fact, it does exactly that {\it below}
the threshold of $\lir \sim \eten{4.5}$ \lsun. This is a clue.
Wu et al. (2005) proposed that there is a ``basic unit" of massive clustered star
formation. For clumps below the mass of the basic unit, the IMF is not fully sampled
and \lir\ increases non-linearly with \mdense. Once the mass exceeds the threshold, 
the IMF is fairly well sampled. Further increases in mass produce more stars but no
further change in \lir/\mdense. In this picture, the difference between star formation
in Galactic molecular clouds, normal galaxies, starbursts, and ULIRGS is simply
how many such basic units (or how much dense gas) they contain. To be concrete,
a \lir\ of \eten{5} \lsun\ corresponds to a stellar mass of about 30 to 40 \lsun.
The basic unit contains 300 to 1000 \msun\ of dense ($n > \eten{5}$ \cmv)
gas. This picture also explains why the scatter in \lir/\lhcn\ is greater in the
Galactic clumps than in galaxies: the most massive star formed will be subject
to stochastic fluctuations, which can produce order of magnitude fluctuations in
the ratio. These tend to average out for a whole galaxy.

Second, we can ask why \lir/\lco\ increases with \lir. While CO traces the
overall mass in Galactic clouds, it fails completely to trace the mass in
dense clumps and cores. This failure is not surprising as CO is optically thick
and thermalized easily. CO does not trace the gas that is relevant to star
formation. Roughly speaking \lhcn/\lco\ provides an estimate of the fraction
of dense gas, as long as that fraction is not too large. Indeed, this ratio
increases with \lir, as expected in this picture \citep{gao2004a}.

\section{Future Directions}

Further studies are needed in both the Galactic and extragalactic context.
Systematic, unbiased surveys of the Galactic plane for dense clumps will
remove the biases in the samples used so far and provide a much larger data set.
Already the Bolocam Galactic Plane Survey has found thousands of clumps 
\citep{williams2007} and deeper surveys are on the way. With large samples in different
environments, we can begin to understand the dependence of the relations on 
chemistry, metallicity, environment, etc. Extragalactic studies should test the
relations at lower \lir, aided by the tremendous sensitivity of {\it ALMA}.
Further studies of the very luminous end will also be important, and higher $J$
lines of dense gas tracers can be studied in many more sources.

We also need to understand the non-linear Kennicutt relations and their
relationship to the linear relations for dense gas. Theories need to explain
{\it both} relations. It is important to realize that the gas surface density,
which appears in the Kennicutt relations, is two steps removed from the actual
star forming entities: dense clumps and cores. We need to understand how
the large-scale surface density controls the formation of molecular clouds
and what controls the formation of dense clumps and cores within those clouds.
The first step may be best studied with high resolution observations of
other galaxies and detailed comparison to simulations with sufficient resolution
to separate individual molecular clouds. The second step, from molecular clouds to
dense clumps and cores, will be hard to study in other galaxies because of
resolution issues, though insights may be available from studies of other galaxies
to see how \lhcn/\lco, for example, depends on conditions. For the most part,
though, we need to study the formation of dense cores from molecular clouds
with more unbiased and complete studies of molecular clouds in our Galaxy.

I will close by thanking the organizers for providing a forum that allowed
some of us touching different parts of the elephant to struggle at least toward
a common language.


\acknowledgements 
I am grateful to Jingwen Wu, whose work underlies the latter part of this paper. Many of
my c2d colleagues have provided information in advance of publication, most
notably Melissa Enoch. Leo Blitz provided guidance on local surface densities of gas.
This work has been supported by NSF Grants AST-0307250 and AST-0607793. 
Additional support came from NASA Origins grants NNG04GG24G and NNX07AJ72G.
The c2d project was part of the
\spitzer\ Legacy Science Program, with support provided by NASA through contracts
1224608, 1230779, and 1230782 issued by the Jet Propulsion Laboratory,
California Institute of Technology, under NASA contract 1407.


\begin{thebibliography}{}

\bibitem[Allen et al.(2007)]{allen2007} Allen, L., et al.\ 2007, 
Protostars and Planets V, 361

\bibitem[Alves et al.(2007)]{alves2007} Alves, J., Lombardi, M., 
\& Lada, C.~J.\ 2007, \aap, 462, L17

\bibitem[Benjamin et al.(2003)]{benjamin2003} Benjamin, R.~A., et 
al.\ 2003, \pasp, 115, 953

\bibitem[Beuther et al.(2002)]{beuther2002} Beuther, H., Schilke, 
P., Menten, K.~M., Motte, F., Sridharan, T.~K., \& Wyrowski, F.\ 2002, 
\apj, 566, 945

\bibitem[Blitz \& Rosolowsky(2006)]{blitz2006} Blitz, L., \& 
Rosolowsky, E.\ 2006, \apj, 650, 933

\bibitem[Bonnell \& Bate(2006)]{bonnell2006} Bonnell, I.~A., \& 
Bate, M.~R.\ 2006, \mnras, 370, 488

\bibitem[Clark et al.(2007)]{clark2007} Clark, P.~C., Klessen, 
R.~S., \& Bonnell, I.~A.\ 2007, ArXiv e-prints, 704, arXiv:0704.2837

\bibitem[Clemens et al.(1988)]{clemens1988} Clemens, D.~P., 
Sanders, D.~B., \& Scoville, N.~Z.\ 1988, \apj, 327, 139 

\bibitem[Dame(1993)]{dame1993} Dame, T.~M.\ 1993, Back to the 
Galaxy, 278, 267

\bibitem[Elmegreen \& Scalo(2004)]{elmegreen2004} Elmegreen, B.~G., 
\& Scalo, J.\ 2004, \araa, 42, 211

\bibitem[Enoch et al.(2006)]{enoch2006} Enoch, M.~L., et al.\ 
2006, \apj, 638, 293

\bibitem[Enoch et al.(2007)]{enoch2007} Enoch, M.~L., Glenn, J., 
Evans, N.~J., II, Sargent, A.~I., Young, K.~E., \& Huard, T.~L.\ 2007, 
ArXiv e-prints, 705, arXiv:0705.3984

\bibitem[Evans et al.(2003)]{2003PASP..115..965E} Evans, N.~J., et al.\
2003, \pasp, 115, 965

\bibitem[Gao \& Solomon (2004a)]{gao2004a}
Gao, Y., \& Solomon, P.\ M., 2004a, \apj, 606, 271

\bibitem[Gao \& Solomon (2004b)]{gao2004b}
Gao, Y., \& Solomon, P.\ M., 2004b, \apjs, 152, 63

\bibitem[Graci{\'a}-Carpio et al.(2006)]{carpio2006} 
Graci{\'a}-Carpio, J., Garc{\'{\i}}a-Burillo, S., Planesas, P., \& Colina, 
L.\ 2006, \apjl, 640, L135

\bibitem[Greve et al. (2005)]{greve2005}
Greve, T.\ R., Bertoldi, F., Smail, Ian, Neri, R., Chapman, S.\ C., Blain, 
A.\ W., Ivison, R.\ J., Genzel, R., Omont, A., Cox, P., Tacconi, L., 
Kneib, J. -P. 2005, \mnras, 359, 1165. 

\bibitem[Greve et al.(2006)]{greve2006} Greve, T.~R., 
Papadopoulos, P.~P., Gao, Y., \& Radford, S.~J.~E.\ 2006, ArXiv 
Astrophysics e-prints, arXiv:astro-ph/0610378 

\bibitem[Harvey et al.(2007)]{harvey2007} Harvey, P.~M., et al.\ 
2007, ArXiv e-prints, 704, arXiv:0704.0253

\bibitem[Hillenbrand(1997)]{hillenbrand1997} Hillenbrand, L.~A.\ 1997, 
\aj, 113, 1733

\bibitem[Hirota et al.(2007)]{hirota2007} Hirota, T., et al.\ 
2007, ArXiv e-prints, 705, arXiv:0705.3792

\bibitem[Kennicutt(1998)]{kennicutt1998} Kennicutt, R.~C., Jr.\ 1998, 
\araa, 36, 189

\bibitem[Krumholz \& McKee(2005)]{krumholzmckee2005} Krumholz, M.~R., \& 
McKee, C.~F.\ 2005, \apj, 630, 250

\bibitem[Krumholz et al.(2005)]{krumholz2005} Krumholz, M.~R., 
McKee, C.~F., \& Klein, R.~I.\ 2005, \nat, 438, 332

\bibitem[Krumholz \& Tan(2007)]{krumholztan2007} Krumholz, M.~R., \& 
Tan, J.~C.\ 2007, \apj, 654, 304

\bibitem[Krumholz \& Thompson(2007)]{krumholzthompson2007} Krumholz, M.~R., 
\& Thompson, T.~A.\ 2007, ArXiv e-prints, 704, arXiv:0704.0792 

\bibitem[Lada \& Lada (2003)]{lada2003}
Lada, C.\ J. \& Lada, E.\ A. 2003,  Ann. Rev. Ast. \& Astrophys., 41, 57

\bibitem[Levine et al.(2006)]{levine2006} Levine, E.~S., Blitz, 
L., \& Heiles, C.\ 2006, \apj, 643, 881

\bibitem[Martel et al.(2006)]{martel2006} Martel, H., Evans, 
N.~J., II, \& Shapiro, P.~R.\ 2006, \apjs, 163, 122

\bibitem[Massey et al.(1995)]{massey1995} Massey, P., Johnson, 
K.~E., \& Degioia-Eastwood, K.\ 1995, \apj, 454, 151

\bibitem[McKee \& Williams(1997)]{mckee1997} McKee, C.~F., \& 
Williams, J.~P.\ 1997, \apj, 476, 144

\bibitem[Mueller et al.(2002)]{mueller2002} Mueller, K.~E., 
Shirley, Y.~L., Evans, N.~J., II, \& Jacobson, H.~R.\ 2002, \apjs, 143, 469

\bibitem[Paglione et al.(1997)]{paglione1997} Paglione, T.~A.~D., 
Jackson, J.~M., \& Ishizuki, S.\ 1997, \apj, 484, 656

\bibitem[Papadopoulos(2007)]{papadopoulos2007} Papadopoulos, P.~P.\ 
2007, \apj, 656, 792

\bibitem[Plume et al. (1992)]{plume1992}
Plume, R., Jaffe, D.\ T., Evans, N.\ J.\ II. 1992, \apjs, 78, 505

\bibitem[Plume et al. (1997)]{plume1997}
Plume, R., Jaffe, D.\ T., Evans, N.\ J.\ II, Martin-Pintado, J., \&
Gomez-Gonzalez, J. 1997, \apj, 476, 730

\bibitem[Reipurth et al.(2007)]{ppv} Reipurth, B., Jewitt, 
D., \& Keil, K.\ 2007, Protostars and Planets V

\bibitem[Ridge et al.(2006)]{ridge2006} Ridge, N.~A., et al.\ 
2006, \aj, 131, 2921

\bibitem[Schmidt (1959)]{Schmidt (1959)}
Schmidt, M. 1959, \apj, 129, 243

\bibitem[Shirley et al.(2003)]{shirley2003} Shirley, Y.~L., Evans, 
N.~J., II, Young, K.~E., Knez, C., \& Jaffe, D.~T.\ 2003, \apjs, 149, 375 

\bibitem[Shu et al. (1987)]{shu1987} 
Shu, F.\ H., Adams, F.\ C., and Lizano, S. 1987, \apj 1987,
Ann. Rev. Ast. \& Astrophys., 25, 23

\bibitem[Shu et al.(2007)]{shu2007} Shu, F.~H., Allen, R.~J., 
Lizano, S., \& Galli, D.\ 2007, \apjl, 662, L75

\bibitem[Solomon \& Vanden Bout (2005)]{Solomon and Vanden Bout (2005)}
Solomon., P.\ M. \& Vanden Bout, P.\ A. 2005, Ann. Rev. Ast.
\& Astrophys., 43, 677

\bibitem[Sridharan et al.(2002)]{sridharen2002} Sridharan, T.~K., 
Beuther, H., Schilke, P., Menten, K.~M., \& Wyrowski, F.\ 2002, \apj, 566, 
931

\bibitem[Williams et al.(2007)]{williams2007} Williams, J.~P., et 
al.\ 2007, American Astronomical Society Meeting Abstracts, 210, \#12.11

\bibitem[Wu et al.(2005)]{wu2005} Wu, J., Evans, N.~J., II, 
Gao, Y., Solomon, P.~M., Shirley, Y.~L., \& Vanden Bout, P.~A.\ 2005, 
\apjl, 635, L173 

\bibitem[Wu \& Evans (2003)]{wu2003}
Wu, J., \& Evans, N.\ J.\ II, 2003, \apjl, 592, 79

\bibitem[Young et al.(2006)]{young2006} Young, K.~E., et al.\
2006, \apj, 644, 326

\bibitem[Zuckerman \& Evans(1974)]{zuckerman1974} Zuckerman, B., \& 
Evans, N.~J., II 1974, \apjl, 192, L149

\end{thebibliography}
\end{document}